\documentclass[
	a4paper,
	10pt,
	unnumberedsections, 
	twoside, 
]{LTJournalArticle}
\usepackage[dvipsnames]{xcolor}
\usepackage{lineno,hyperref,xcolor,float,caption,paralist,siunitx,amssymb} 
\usepackage[singlespacing]{setspace}
\usepackage{soul} 
\modulolinenumbers[5]
\usepackage[T1]{fontenc}
\usepackage{wrapfig}
\usepackage{multicol}
\usepackage{changes}

\runninghead{} 

\footertext{}

\title{A multi-reflection time-of-flight mass spectrometer for the offline ion source of the PUMA experiment}

\author{
	M. Schlaich\textsuperscript{a}\thanks{Corresponding author: \href{mailto:mschlaich@ikp.tu-darmstadt.de}{mschlaich@ikp.tu-darmstadt.de}}, J. Fischer\textsuperscript{a}, P. Fischer\textsuperscript{b}, C. Klink\textsuperscript{a}, A. Obertelli\textsuperscript{a}, A. Schmidt\textsuperscript{a}\\L. Schweikhard\textsuperscript{b} and F. Wienholtz\textsuperscript{a}
}

\date{\footnotesize\textsuperscript{\textbf{a}}Technische Universität Darmstadt, Institut für Kernphysik, Schloßgartenstraße 9, 64289 Darmstadt, Germany\\ \textsuperscript{\textbf{b}}Universität Greifswald, Institut für Physik, Felix-Hausdorff-Straße 6, 17489 Greifswald, Germany}

\renewcommand{\maketitlehookd}{
	\begin{abstract}
		\noindent The antiProton Unstable Matter Annihilation experiment (PUMA) at CERN aims at investigating the nucleon composition in the matter density tail of radioactive as well as stable isotopes by use of low-energy antiproton-nucleon annihilation processes. For this purpose, antiprotons provided by the Extra Low ENergy Antiproton (ELENA) facility will be trapped together with the ions of interest. While exotic ions will be obtained by the Isotope mass Separator On-Line DEvice (ISOLDE), stable ions will be delivered from an offline ion source setup designed for this purpose. This allows the proposed technique to be applied to a variety of stable nuclei and for reference measurements. For beam purification, the ion source setup includes a multi-reflection time-of-flight mass spectrometer (MR-ToF MS). Supported by SIMION\textsuperscript{\textregistered} simulations, an earlier MR-ToF MS design has been modified to meet the requirements of PUMA. During commissioning of the new MR-ToF device with Ar$^+$ ions, mass resolving powers in excess of 50,000 have been obtained after 150 revolutions, limited by the chopping of the continuous beam from an electron impact ionisation source.  
	\end{abstract}
}

\begin{document}
\maketitle 
\section{Introduction}

To date, the many-body problem of nuclear physics cannot be solved exactly and approximations are required \cite{Hergert2020}. To contribute to the understanding of nuclear structure, radioactive ion beam facilities provide the investigation of a broad variety of short-lived nuclei which allow to explore nuclear systems beyond isospin symmetry. On the neutron rich side of the nuclear landscape, nuclei form neutron skins, referring to an excess of neutrons on the nuclear surface. Its size, the neutron skin thickness, can be determined by various methods \cite{Zenihiro2010,KRASZNAHORKAY2004224,Klos2007,Adhikari2021,Tarbert2014} and is correlated to the nuclear equation of state through the slope of the symmetry energy, \textit{i.e.} the contribution of the proton-to-neutron asymmetry in nuclear matter \cite{Roca-Maza2011}. Bridging the gap between the smallest and largest nucleonic systems known, this correlation is also of relevance for the structure of neutron stars \cite{Horowitz-neutronstar, Hebeler_neutronstars}. Proceeding to even more asymmetric systems than neutron-skin nuclei, halo nuclei located at the nuclear drip lines \cite{Tanihata1985.55.2676} offer the possibility of studying nuclear few-body systems \cite{FREDERICO2012939}. These are nuclei in which one or more nucleons have an increased probability density within classically forbidden regions, far beyond the nuclear core potential \cite{Hansen1995,Riisager_2013}.
\\
To contribute to the understanding of phenomena that occur beyond the half-density radius of nuclei, such as neutron skins and halo nuclei, the antiProton Unstable Matter Annihilation (PUMA) experiment will investigate the nucleon composition at the tail of the matter density distribution of stable as well as radioactive isotopes using low-energy antiprotons \cite{PUMA, Zacarias2022}. By trapping both matter and antimatter together and studying the antiproton-nucleon annihilation residuals, which are mainly pions, the PUMA experiment aims to determine the proton-to-neutron ratio in the tail of the nuclear matter density distribution. However, there is to date no facility worldwide that provides both radioactive nuclei and low-energy antiprotons at the same location. The PUMA experiment therefore plans to transport the antiprotons within the CERN site from the Extra Low ENergy Antiproton (ELENA) facility \cite{Maury2014}, located at the Antimatter Factory, to the Isotope mass Separator On-Line DEvice (ISOLDE) \cite{Kugler_2000,Catherall_2017} in a transportable Penning trap.
\\
In order to start with stable-ion experiments, PUMA requires mass-separated and cooled ion bunches that can be obtained from a broad mass range of isotopes. Therefore, a versatile offline ion source system is currently under development at the Technical University of Darmstadt. Following the ion source, a multi-reflection time-of-flight mass spectrometer (MR-ToF MS) \cite{WOLLNIK1990267,WOLLNIK201338} for beam purification and a radio-frequency quadrupole (RFQ) for particle accumulation and beam cooling \cite{HERFURTH2001254,Nieminen2022,Mane2009,VALVERDE2020330} will be used. State-of-the-art MR-ToF systems provide mass resolving powers beyond $10^5$ \cite{FISCHER2021,DICKEL2015172}, reached within milliseconds \cite{ROSENBUSCH2023167824}. As such, they are the tool of choice for fast, high-precision ion beam purification.
\begin{figure*}[ht]
	\centering
	\includegraphics[width = 1.0\textwidth]{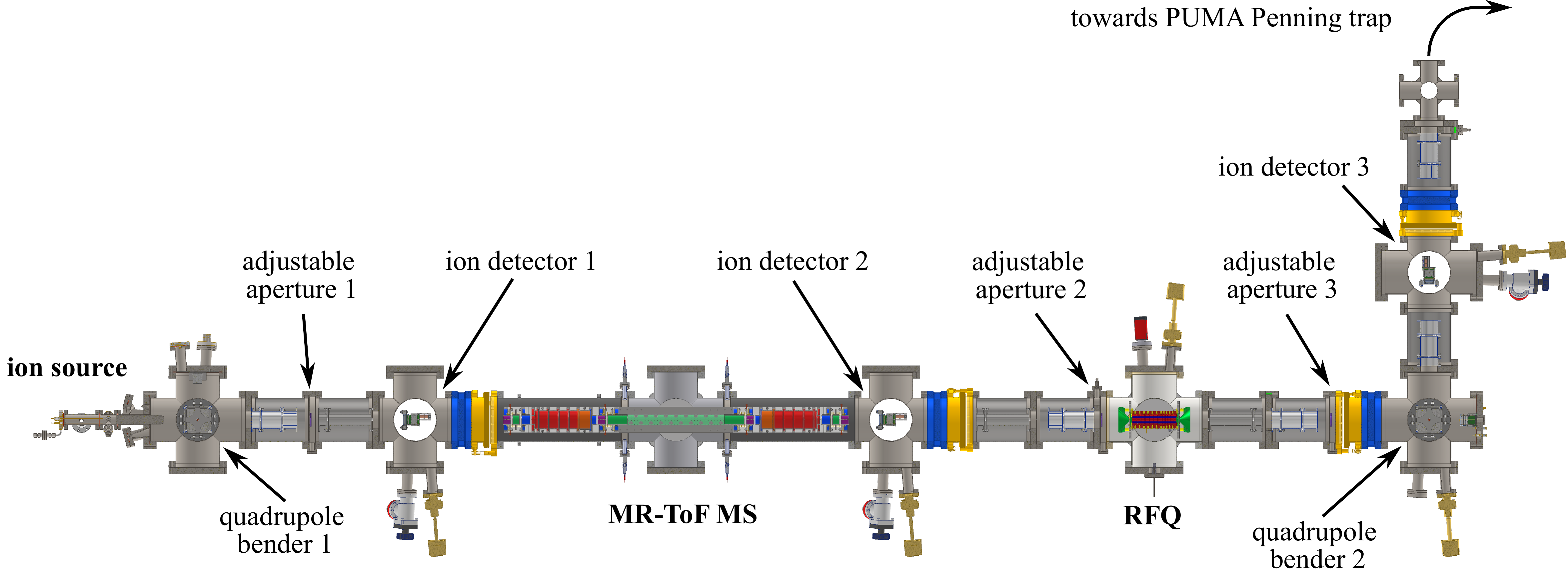}
	\caption{CAD representation of the PUMA offline ion source beamline. The ions generated in the ion source are mass-separated in the MR-ToF MS and cooled and bunched in the RFQ. Three movable time-of-flight (ToF) detectors allow a stepwise characterisation of the ion beam.}
	\label{Fig:Offline_ion_source}
\end{figure*}
\noindent

\section{The PUMA offline ion source beamline}\label{Sec:Offline_ion_source}
The offline ion source beamline can be divided into three main sections which - in beam direction - are the ion source, the MR-ToF MS and the RFQ (see Fig. \ref{Fig:Offline_ion_source}). They are followed by a fourth section to gradually improve the vacuum behind a 90-degree bend towards the PUMA beamline. Furthermore, the offline ion source is complemented with three movable ion detection units, three iris-type apertures for adjustments of the gas flow and ion optical components for focusing, steering and energy modification (not labelled in Fig. \ref{Fig:Offline_ion_source}). The offline setup is to be installed at the PUMA experimental area located at the ELENA facility.
\subsection{Overview of the setup}
In a commercial electron impact ionisation source (IQE 12/38, SPECS Surface Nano Analysis GmbH, Berlin, Germany), a gaseous material is ionised and the ions are extracted continuously with beam currents of a few $\SI{}{\pico\ampere}$ up to $\SI{800}{\nano\ampere}$ (maximum current specified by manufacturer for argon ion-beam production). The ion source is equipped with two pairs of deflection plates (\SI{6}{\milli\meter} $\times$ \SI{16}{\milli\meter}), each \SI{5}{\milli\meter} spaced, which are aligned perpendicular to each other. When switching the electric potentials applied to the plates between an ion blocking and transmission state, chopped ion packets can be extracted, which is required for the operation of the MR-ToF MS. Bunches with a full width of half maximum (FWHM) minimum of about \SI{250}{\nano\second} can be extracted. The ion source is attached to a quadrupole ion bender, which provides the possibility to add other ion source types at a later stage. Downstream, the MR-ToF MS allows beam analysis and mass-selective ion ejection towards the RFQ, in which the ions remain trapped until up to $10^5$ ions are accumulated in one bunch. While ion accumulation is the main purpose of the RFQ, the ion bunch emittance can simultaneously be reduced within milliseconds through buffer gas cooling. Subsequently, the ions are ejected and guided towards the PUMA Penning trap, in which they annihilate with the trapped antiprotons.

\subsection{The MR-ToF mass spectrometer}\label{Sec:Starting point}
MR-ToF systems, operated in various nuclear laboratories around the world \cite{WOLF2013123,ROSENBUSCH2020184,DICKEL2015172,Jesch2015,HIRSH2016229}, share the operating principle that charged particles are trapped between two opposing ion-optical mirrors and ejected after up to thousands of revolutions. In this way, the ion flight path can reach several kilometres in experimental setups that measure only a few meters. Furthermore, the reflecting potentials are tuned to retain small temporal bunch widths by compensating flight-time differences resulting from small discrepancies in the ions' energies. Ion species differing in mass-over-charge ratio and accelerated by the same electric potential can thus be separated with isobaric precision. The two electrostatic mirrors are usually made of stacks of cylindrical electrodes for better control of the potential gradients and are separated by a field-free drift region \cite{ISHIDA2004468}. 
\\
The design of the PUMA MR-ToF MS is a further development that started from the present MR-ToF setup at the University of Greifswald \cite{FISCHER2021,KNAUER2019116189, Fischer2019.1.033050,Fischer2020.2.043177}, which includes lens and steering optics in front of the entrance and behind the exit mirror, in-trap deflector electrodes for in-trap separation of unwanted ion species \cite{Fischer2018015114} and an in-trap lift electrode. The latter can be used to capture ions at adjustable storing energies as well as to eject a desired ion species after mass separation \cite{WOLF20128,Wienholtz2017, KNAUER201746}. The PUMA offline ion source is intended to allow for fast accumulation of up to $10^5$ mass-selected ions inside the RFQ before they are sent to the antiprotons in the PUMA Penning trap to obtain annihilation event rates that can be distinguished from background annihilations. Therefore, the design aims to maximise the transversal acceptance of the MR-ToF device to allow for a large ion throughput while guaranteeing a sufficient mass resolving power of up to 10$^5$. 
\\\\
Intuitively, the acceptance is expected to increase with the inner diameter of the electrodes. To our knowledge, however, a correlation is not known and, thus, it is investigated in the following with simulations. The maximum possible electrode size is limited by the vacuum tubes, which are chosen to have an inner diameter of \SI{155}{\milli\meter} in the present design. Due to limited space of the experimental site at ELENA, it is decided to allow a maximum length of \SI{1.2}{\meter} for the MR-ToF MS. In analogy to the Greifswald design, the MR-ToF system should include injection and ejection ion optics, the in-trap deflector and lift electrodes. However, it is decided not to include the support for temperature compensation \cite{KNAUER2019116189} in the new MR-ToF system, as it has been found that the stability of the potentials is much more important than the influence of temperature fluctuations \cite{Wienholtz2020}. This allows a simpler and more modular design. Furthermore, similar to the Greifswald design, it is decided to use six electrodes per ion-mirror stack, including one electrode on negative potential which is intended to form lens-like potentials for radial refocusing and five electrodes on positive potential. The operation of existing MR-ToF systems has shown that this provides sufficient degrees of freedom to shape the potential profile to reach mass resolving powers beyond 100,000.

\section{Simulations}\label{Sec:Simulation}
For the optimisation of the electrode geometry, simulations are performed using SIMION\textsuperscript{\textregistered} \cite{DAHL20003}. This allows to define customised electrode geometries to which a potential can be assigned. The software then calculates the resulting electric fields by solving the Laplace equation. Based on the potential array thus determined, ion trajectories are calculated as a function of the starting values of the ions. In all simulations described, we use cations. The application to anions requires a polarity change of the specified potential values.

\subsection{Simulation method}\label{Sec:simulation-method}
Using SIMION\textsuperscript{\textregistered}, the ion-optical acceptance of a given electrode geometry is determined by measuring the emittance of the ions that successfully pass through the MR-ToF system. However, this strongly depends on the six-dimensional potential set which is applied to the mirror electrodes. Here, we assume that the same potential is applied mirror-symmetrically to opposing electrodes. A potential combination that allows sufficiently isochronous ion reflection (time dispersion of the ion bunch per revolution in the order of a few tens of picoseconds), again depends on the electrode geometry and can only be found with a more time-consuming optimisation algorithm. However, this approach can result in local minima, which means that comparability between different geometries is not necessarily given. Therefore, an alternative measure is used that indirectly reflects the ion acceptance and is comparable for different designs, regardless of the electrode size and shape: We use the volume of the subspace formed by all mirror potential sets $V\in\mathbb{R}^6$ that allow a given ion bunch to be trapped without losses. If the initial conditions for the ions are not changed for the trajectory simulations of different geometries, the solution volume is expected to reflect the acceptance depending on the proposed geometry modifications. Since it is not possible to calculate the described volume analytically, its size is determined using a Monte Carlo (MC) approach.
\\
The voltages that can be applied to the mirror electrodes are limited by the maximum operation voltage of \SI{5}{\kilo\volt} for standard SHV connectors. Furthermore, we restrict the outer five mirror electrodes to a repulsive potential, while the innermost electrode is used as lens with a negative potential applied to it. This provides a finite sample volume $\Omega_\text{m}\subset\mathbb{R}^6$ of voltage combinations for the mirror electrodes. The used mirror potentials $V_\text{i}\in\Omega_\text{m}$ are taken randomly from the defined voltage interval. With the weight function $P(V_\text{i}) \in \{0,1\}$, which is equal to 1 if the applied potential profile allows to trap all ions on loss-free trajectories and equal to 0 if this is not the case, the MC simulation approaches the desired relative solution volume
\begin{equation}\label{Eq:V_rel}
    V_\text{rel} = \frac{\int_{\textbf{V}\in\Omega_\text{m}}P(\textbf{V})d\Omega_\text{m}}{\int_{\textbf{V}\in\Omega_\text{m}}d\Omega_\text{m}}
\end{equation}
by calculating the mean value
\begin{equation}\label{Eq:R_rel_MC}
    \langle V_\text{rel}\rangle = \frac{1}{N} \sum_{i=1}^N P(V_i).
\end{equation}
Here, $N$ is the number of voltage configurations tested. An illustration of the potential volume determined in this kind of MC simulation is provided in Sec. \ref{Sec:Potential_determination}. Unless stated otherwise, 100 $^{85}$Rb$^+$ ions trapped for 20 revolutions are used in each of the following simulations.

\begin{table*}[ht]
    \centering
    \caption{Dimensions of the six mirror electrodes for two different mirror concepts to be compared using the MC simulation. The sizes are chosen to sum up to the same total length of the electrode stack. The labels M1 - M6 correspond to the mirror electrodes from inside out (see Fig. \ref{Fig:final_design}). The spacing between the electrodes remains unchanged.}
    \begin{tabular}{l c c c c c c c}
		\hline 
		mirror concept &  $l_\text{M1}$ & $l_\text{M2}$ & $l_\text{M3}$ & $l_\text{M4}$ & $l_\text{M5}$ & $l_\text{M6}$ & spacing\\
		\hline
		\hline 
		Greifswald MR-ToF analyser & $\SI{40}{\milli\meter}$ & $\SI{30}{\milli\meter}$ & $\SI{20}{\milli\meter}$ & $\SI{12}{\milli\meter}$ & $\SI{12}{\milli\meter}$ & $\SI{12}{\milli\meter}$ & $\SI{5}{\milli\meter}$\\ 
		unified mirror electrodes & $\SI{42}{\milli\meter}$ & $\SI{18}{\milli\meter}$ & $\SI{18}{\milli\meter}$ & $\SI{18}{\milli\meter}$ & $\SI{18}{\milli\meter}$ & $\SI{12}{\milli\meter}$ & $\SI{5}{\milli\meter}$\\ 
		\hline 
    \end{tabular}
    \label{Tab:Dimensions_equal_ME_test}
\end{table*}
\begin{table*}[ht]
    \centering
    \caption{Results of the relative solution space $V_\text{rel}$ in percent of the simulation varying the mirror electrode length $l_\text{M}$ with a fixed inner diameter $d_\text{M} = \SI{68}{\milli\meter}$.}
    \begin{tabular}{c c c c c c}
		\hline 
		$l_\text{M}$ & $\SI{15}{\milli\meter}$ & $\SI{20}{\milli\meter}$ & $\SI{25}{\milli\meter}$ & $\SI{30}{\milli\meter}$ & $\SI{35}{\milli\meter}$ \\
		\hline
		\hline 
		$V_\text{rel}$ & $\SI{5.1\pm0.2}{\percent}$ & $\SI{7.9\pm0.3}{\percent}$ & $\SI{8.8\pm0.3}{\percent}$ & $\SI{10.8\pm0.3}{\percent}$ & $\SI{10.4\pm0.3}{\percent}$ \\ 
		\hline 
    \end{tabular}
    \label{Tab:Final_ME_optimization}
\end{table*}

\subsection{Determination of the mirror electrode size}
To converge to a final electrode design, the MC simulation approach is applied to MR-ToF systems with varying electrode dimensions. The design currently in use at the University of Greifswald has mirror electrodes of different lengths. As a first step, it is investigated whether the unification of the mirror electrode length has an effect on the solution volume. Identical mirror electrodes provide the advantages to make use of mass production. Therefore, the MC simulation is applied to two different mirror concepts with electrode lengths as specified in Tab. \ref{Tab:Dimensions_equal_ME_test}. The length of the unified electrodes is chosen such that the total length of the electrostatic mirror equals that of the Greifswald MR-ToF analyser while the length of the negative mirror electrode (M1) is on the same order of magnitude.\\
The ion bunch used for the simulation has a transverse 2$\sigma$ emittance of 3.7$\pi$ \SI{}{\milli\meter~\milli\radian} and a Gaussian energy distribution around $\SI{2000}{\electronvolt}$ with a standard deviation of $\SI{50}{\electronvolt}$. According to the MC simulation, this leads to the solution volumes of $V_\text{rel,u} = \SI{1.2\pm0.03}{\percent}$ for unified electrodes and $V_\text{rel,G} = \SI{1.26\pm0.03}{\percent}$ for the Greifswald design. Considering the uncertainties, the results are compatible with each other. Therefore, the advantages of uniform electrodes are considered to be more important and it is decided to use mirror electrodes of equal length, which has yet to be determined. 
\\
As stated above, it is to be expected that the acceptance of the MR-ToF analyser is increased by increasing the inner diameter of the system, \textit{i.e.} the diameter of the in-trap lift $d_\text{lift}$ and that of the mirror electrodes $d_\text{M}$. Also, their ratio $d_\text{lift}$/$d_\text{M}$ as well as the ratio $l_\text{M}$/$d_\text{M}$, where $l_\text{M}$ is the length of a mirror electrode, may affect the acceptance. 
To investigate the effect of the mirror electrode size, MC simulations are conducted for fifteen different mirror designs with varying electrode lengths (\SI{8}{\milli\meter}, \SI{11}{\milli\meter}, \SI{14}{\milli\meter}, \SI{17}{\milli\meter} and \SI{20}{\milli\meter}) and inner diameters (\SI{18}{\milli\meter}, \SI{30}{\milli\meter} and \SI{50}{\milli\meter}). The values are chosen to meet the mechanical and experimental length restrictions and to feature reasonable dimensional differences in order to see a trend of $V_\text{rel}$ as a function of the electrodes' dimension. Further, the diameter value of \SI{18}{\milli\meter} was chosen to match the mirror diameter to that of the in-trap lift used in this simulation. 
\begin{figure}[h]
	\centering
	\includegraphics[width = 0.5\textwidth]{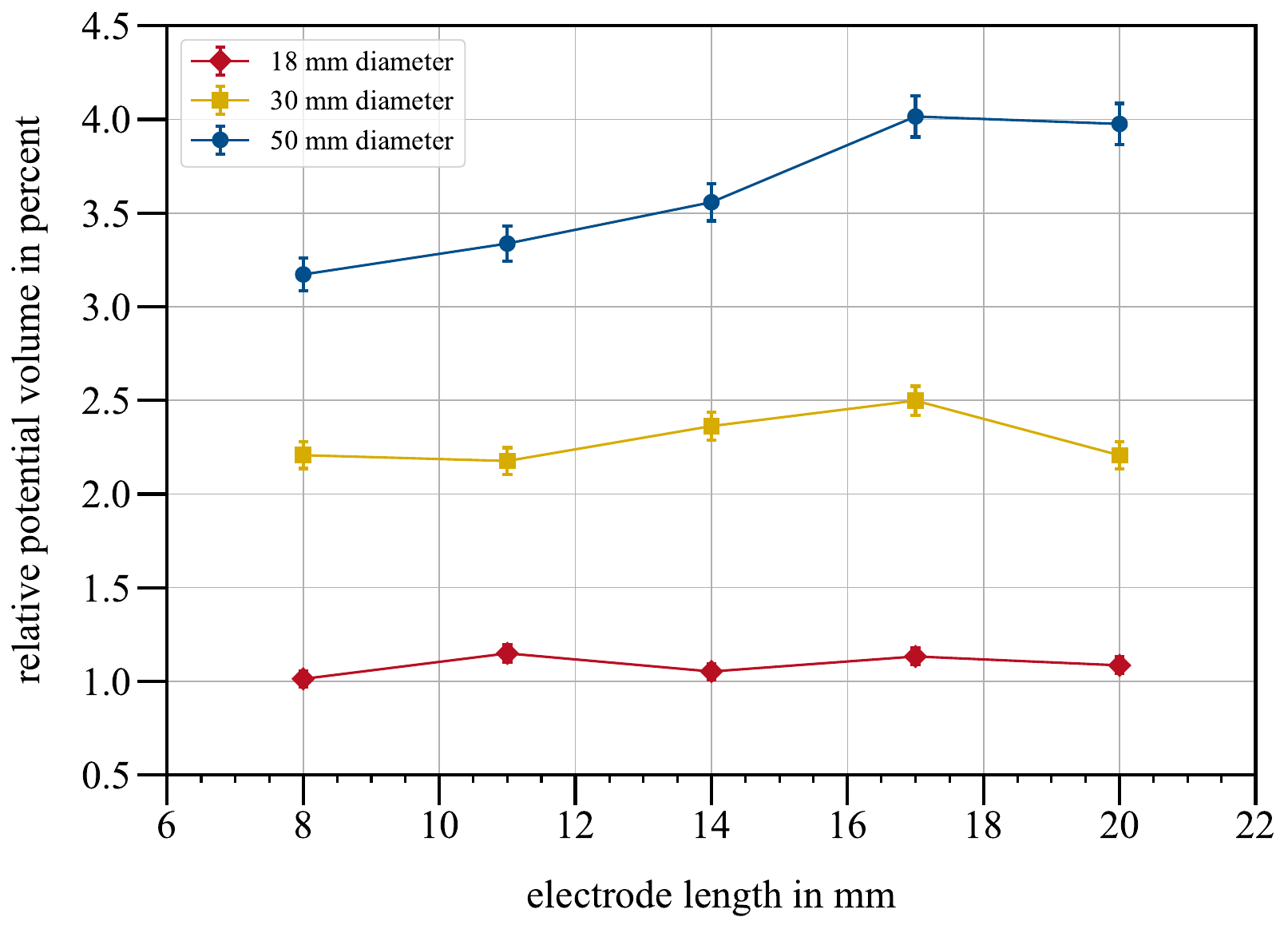}
	\caption{Relative solution space found in the MC simulations as a function of the mirror electrode inner diameter and length. For more details see the text.}
	\label{Fig:ME_design}
\end{figure}
\noindent
The results are given in Fig. \ref{Fig:ME_design}. Confirming the expectations, we see that $V_\text{rel}$ grows with increasing inner diameter, which is interpreted as an increased acceptance. Moreover, it turns out that there is an optimum electrode length as a function of its diameter, which becomes more apparent for larger diameters. Based on these results, we decide to use mirror electrodes with the largest diameter compatible with the current design constraints, which is $d_\text{M}=\SI{68}{\milli\meter}$ and is simulated below.
\\
With the diameter determined, the mirror electrode length is varied, keeping the in-trap lift electrode dimensions at $l_\text{lift}=\SI{600}{\milli\meter}$ and $d_\text{lift}=\SI{18}{\milli\meter}$. As can be seen in Tab. \ref{Tab:Final_ME_optimization}, a maximum $V_\text{rel}$ is found for $l_\text{M}\geq\SI{30}{\milli\meter}$. Larger mirror electrodes are not considered because long electrodes come at the expense of a shorter in-trap lift electrode and thus reduced longitudinal acceptance due to the given limit of the MR-ToF MS length (see Sec. \ref{Sec:Starting point}). The negative mirror electrode (M1) is optimised using the same algorithm leading to a length of $\SI{40}{\milli\meter}$.
\begin{figure*}[h]
	\centering
	\includegraphics[width = 1.0\textwidth]{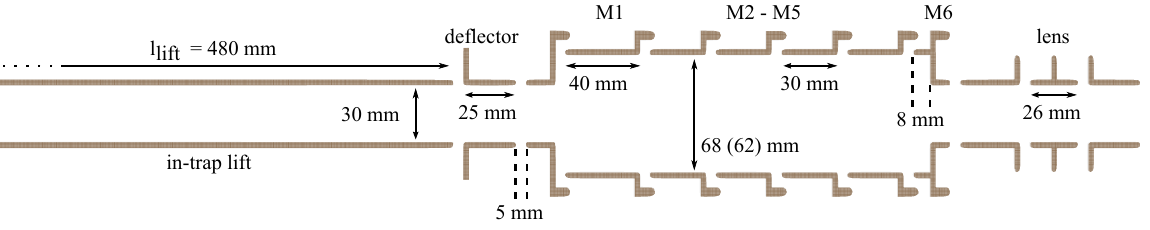}
	\caption{Cross-sectional view of the final MR-ToF MS geometry implementation in SIMION\textsuperscript{\textregistered} including all relevant dimensions. The distance between each electrode is $\SI{5}{\milli\meter}$. Due to mirror symmetry, only one half of the device is shown. For design reasons, the inner diameter of the mirror electrodes later had to be reduced from \SI{68}{\milli\meter} to \SI{62}{\milli\meter}.}
	\label{Fig:final_design}
\end{figure*}
\noindent
\begin{figure*}[ht]
	\centering
	\includegraphics[width = 0.75\textwidth]{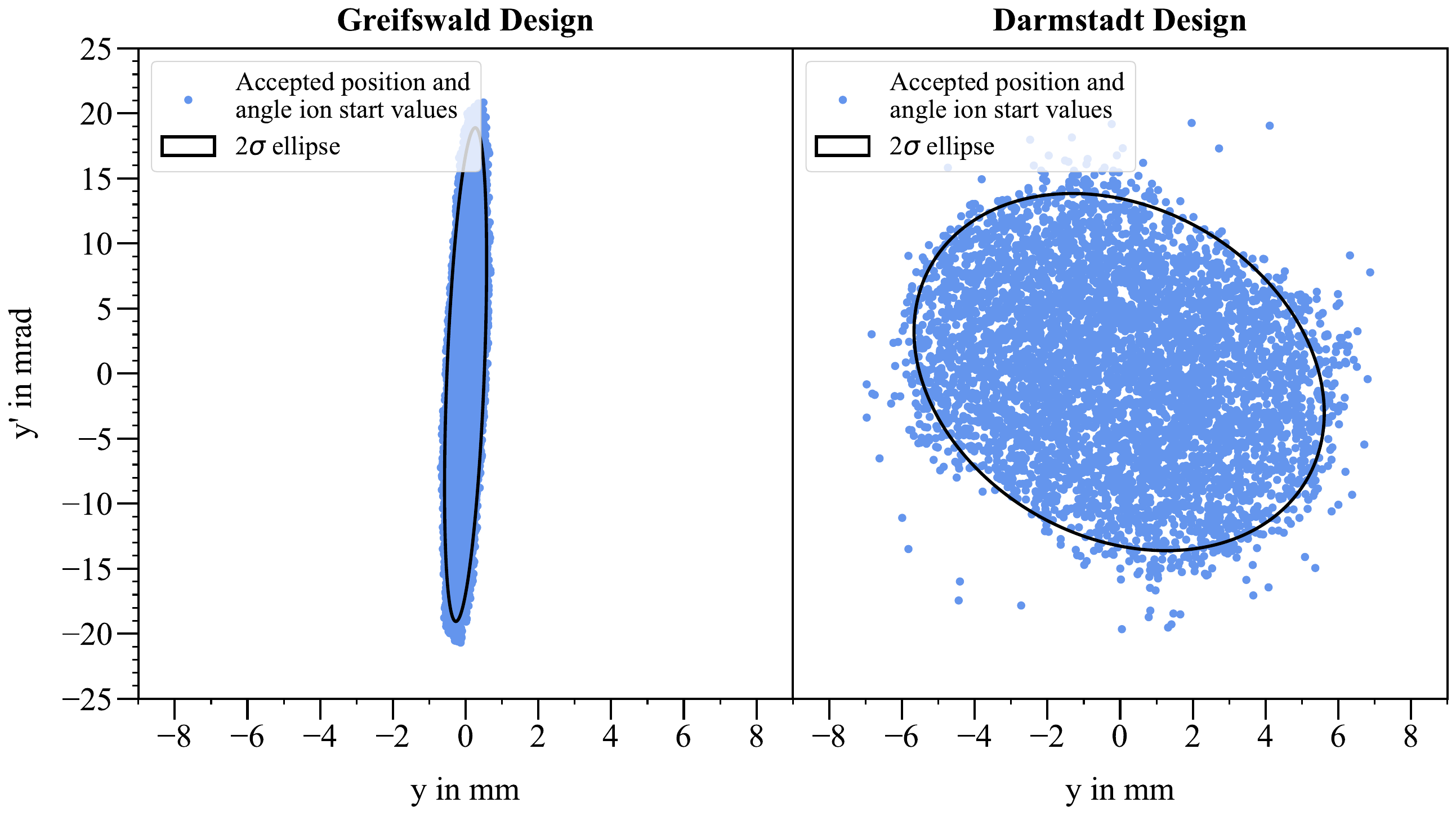}
	\caption{Acceptance plot of the Greifswald (left) and Darmstadt (right) MR-ToF MS geometries. The respective applied potential profiles are selected based on the largest mass resolving power found in the final MC simulation. The second orthogonal part of the phase space, the z-z' plane (beam axis along the x-axis), is redundant due to the rotational symmetry of the electrodes.}
	\label{Fig:Acceptance}
\end{figure*}

\subsection{Final electrode dimensions}
Finalising the geometry of the MR-ToF system, the MC algorithm is used to define the in-trap lift diameter and that of the aperture on the injection and ejection side of the mirror. \SI{30}{\milli\meter} is determined for both dimensions. Furthermore, the length of the in-trap lift, $\SI{480}{\milli\meter}$, is chosen to ensure that the MR-ToF system uses the entire length of the vacuum chamber. A variation of the aperture of mirror electrode M6, which forms the transition between the mirror electrode stack and the lens electrode assembly, shows no relevant influence on the solution volume. It is adjusted to the lift and lens diameter. Due to minor technical difficulties in the design of the mirror electrode assembly, the inner diameter had to be reduced from \SI{68}{\milli\meter} to \SI{62}{\milli\meter}. An overview of the final geometry and relevant dimensions is given in Fig. \ref{Fig:final_design}. It should be noted that all the investigations shown are multi-dimensional problems and the result of each simulation is affected by all electrode dimensions. However, the simulation effort for an absolute optimisation of the MR-ToF MS geometry in dependence of all relevant dimensions on each other is not feasible in terms of simulation cost.
\\
After specifying the electrode geometry, final MC simulations are performed to compare the solution spaces obtained from the Greifswald and the new Darmstadt design. A solution volume of $\SI{1.19\pm0.03}{\percent}$ is found for the Greifswald MR-ToF device and $\SI{6.54\pm0.07}{\percent}$ for the new PUMA MR-ToF system. The solution space is increased by a factor of about six for the considered ion distribution. This result is interpreted as an increased acceptance. To investigate the acceptance change directly, the potential profiles leading to the largest mass resolving power $R = t/\left(2\Delta t\right)$ are selected for each solution space, where $t$ is the mean ToF and $\Delta t$ the FWHM ToF of the simulated ions. Based on these two mirror potential sets and geometries, the starting values of the ions that can be successfully trapped for 100 revolutions and ejected are measured. To include the capture and ejection process, the ions are generated \SI{10}{\milli\meter} in front of the injection lens and stopped \SI{10}{\milli\meter} behind the ejection lens. The result is plotted in Fig. \ref{Fig:Acceptance} together with the 2$\sigma$ ellipse containing $\SI{91}{\percent}$ of the measured data points. Compared to the Greifswald design the area of the 2$\sigma$ ellipse increases from $\SI{9.75\pi}{\milli\meter~\milli\radian}$ to $\SI{75.38\pi}{\milli\meter~\milli\radian}$, directly showing the increase in acceptance by a factor of about eight.
\begin{table*}[ht]
    \centering
    \caption{Set of mirror and in-trap lift potentials optimised for a $\SI{3000}{\electronvolt}$ $^{85}$Rb$^+$ beam.}
    \begin{tabular}{c c c c c c c c}
		\hline 
		electrode & M1 & M2 & M3 & M4 & M5 & M6 & in-trap lift\\
		\hline
		\hline 
		  potential in $\SI{}{\volt}$ & $\SI{-3710}{\volt}$ & $\SI{745}{\volt}$ & $\SI{893}{\volt}$ & $\SI{1393}{\volt}$ & $\SI{2160}{\volt}$ & $\SI{2681}{\volt}$ & $\SI{1003}{\volt}$ \\ 
		\hline 
    \end{tabular}
    \label{Tab:potential_set_commissioning}
\end{table*}
\begin{figure*}[h]
	\centering
	\includegraphics[width = 1.0\textwidth]{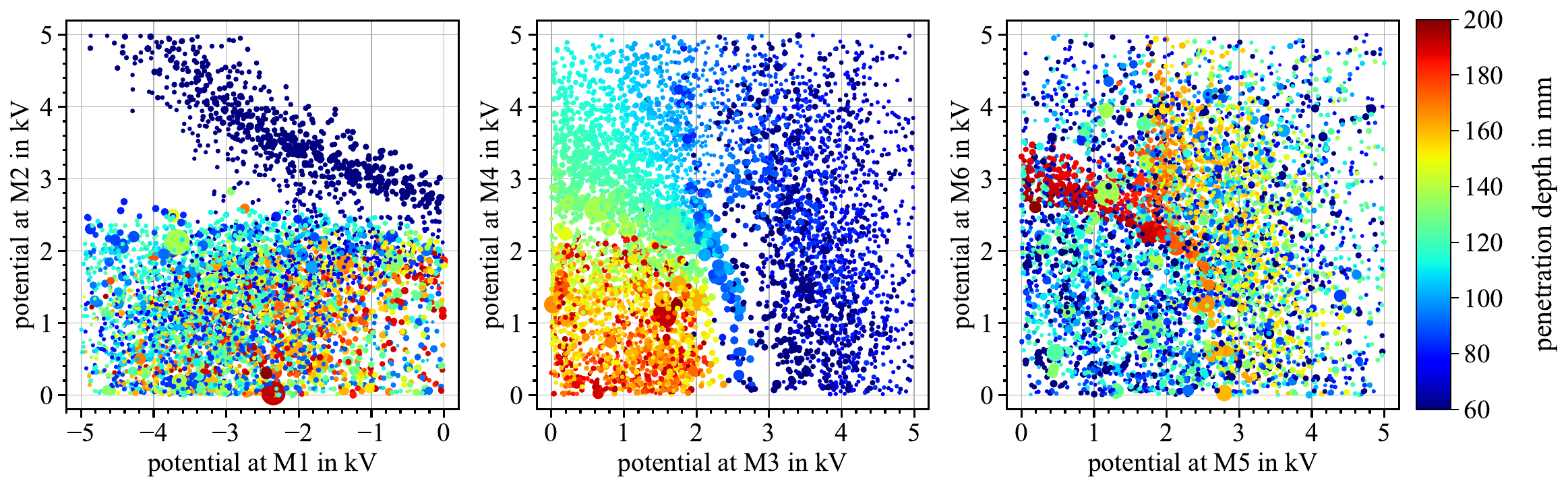}
	\caption{Potential space obtained from the MC simulation described in Sec. \ref{Sec:simulation-method} for the final MR-ToF MS geometry. The spot size indicates the final bunch width of simulated ion bunch and the colour its penetration depth into the electrostatic mirror volume, which has a total length of $\SI{199}{\milli\meter}$.}
	\label{Fig:Pot-space}
\end{figure*}
\noindent
\subsection{Potential profile determination}\label{Sec:Potential_determination}
The MC simulation approach described is not only used to improve the system's acceptance, but also to determine suitable potential profiles. Figure \ref{Fig:Pot-space} illustrates the potential configurations obtained from the MC simulation performed for the final electrode design. While the spot size reflects the value of the final bunch width, the larger the plotted circle the larger the width of the ToF distribution, the colour indicates the penetration depth of the ions into the \SI{199}{\milli\meter} long electrostatic mirror. 
Highlighted by clusters of points of the same colour, subspaces of the potential volume can be identified that allow ion storage at different penetration depths. Stable trajectories are found that have reversal points over almost the entire span of the mirror volume. A penetration depth of \SI{60}{\milli\meter} corresponds to the first \SI{10}{\milli\meter} of the mirror electrode M2, while a depths of \SI{199}{\milli\meter} forms the end of the mirror electrode volume. In order to include all mirror electrodes to contribute to the reflecting mirror potential, voltage configurations that allow a penetration depth of about 160 - \SI{200}{\milli\meter} are considered. In this way, the available six degrees of freedom are exploited, which is expected to be beneficial for the compensation of the time dispersion of the trapped ion bunches. By additionally separating those that lead to the smallest possible bunch widths, suitable mirror potentials can be determined.
\\
For the further improvement of the mirror potential shape, a simplex optimisation routine (Nelder-Mead downhill type) is used, included in a module provided by SIMION\textsuperscript{\textregistered}. The algorithm varies the potentials applied to the six mirror electrodes while locally minimising a user-specific metric. This allows to reduce the final bunch width further and thus increase the mass resolving power after a given number of revolutions.
Based on this approach, the mirror potential set to be used for commissioning measurements is determined for a $\SI{3000}{\electronvolt}$ $^{85}$Rb$^+$ beam captured with a trapping energy of about $\SI{2000}{\electronvolt}$ inside the MR-ToF MS. The relevant potentials are provided in Tab. \ref{Tab:potential_set_commissioning}. While the injection and ejection lens are set to $U_\text{lens}=\SI{2800}{\volt}$, all steering and in-trap deflector segments are kept at $\SI{0}{\volt}$. 
With this specific potential configuration, the mass resolving power achievable is investigated varying the initial ion distribution. Keeping the transverse 2$\sigma$ emittance constant at 8$\pi$ \SI{}{\milli\meter~\milli\radian}, initial bunch widths $\Delta t_\text{i}$ of $\SI{10}{\nano\second}$, $\SI{50}{\nano\second}$ and $\SI{100}{\nano\second}$ (FWHM) are combined with energy distributions $\Delta E_\text{i}$ of $\SI{1}{\electronvolt}$, $\SI{2}{\electronvolt}$ and $\SI{3}{\electronvolt}$ (FWHM). In the ideal case of $\Delta t_\text{i}=\SI{10}{\nano\second}$ and $\Delta E_\text{i}=\SI{1}{\electronvolt}$, the mass resolving power reaches a value of about 630,000 after 2000 revolutions. In contrast, at $\Delta t_\text{i}=\SI{100}{\nano\second}$ and $\Delta E_\text{i}=\SI{3}{\electronvolt}$ the mass resolving power reaches a value of about 230,000 for similar storing times. Although the resolving power for ion selection is generally smaller, the values obtained are sufficient to achieve the objectives of the offline ion source to provide isotopically pure stable ion bunches.

\section{Technical design}
\begin{figure*}[h]
	\centering
	\includegraphics[width = 1.0\textwidth]{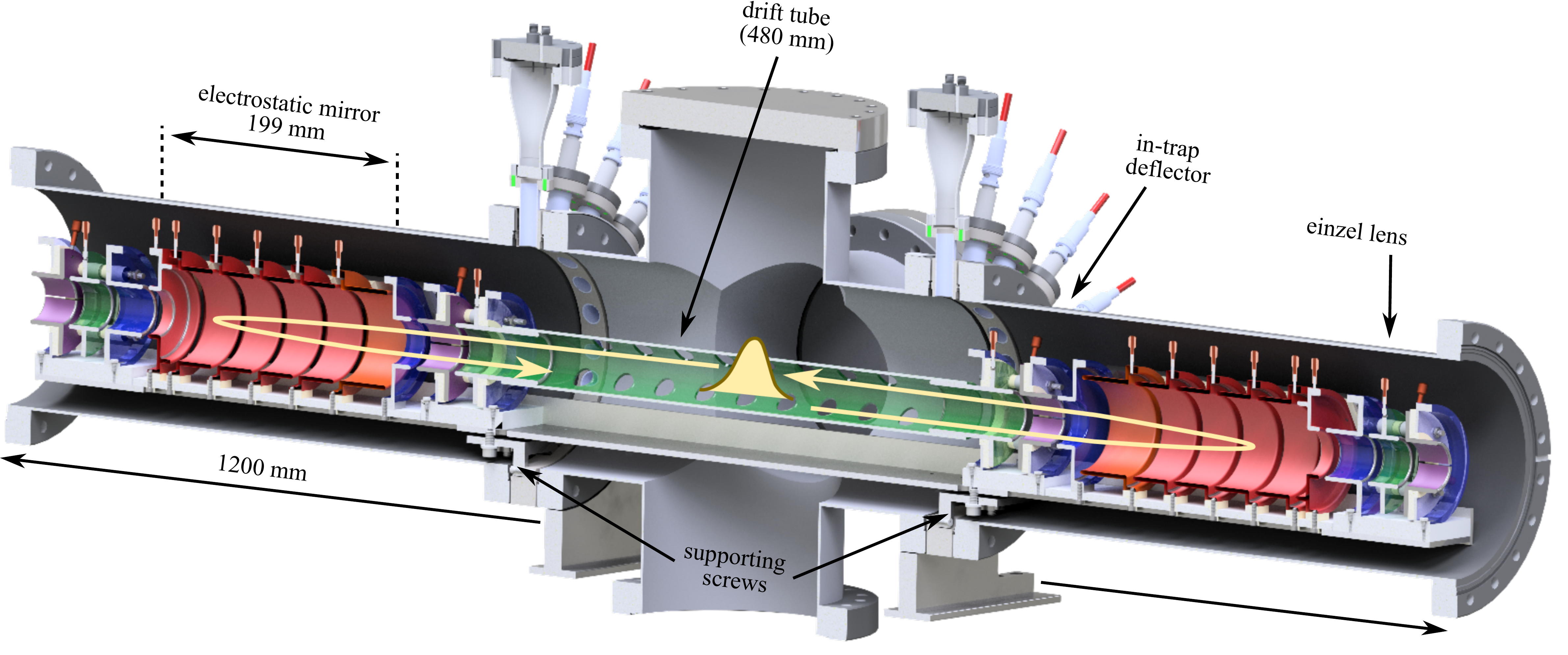}
	\caption{CAD representation in half-section view of the MR-ToF MS. Ions are trapped between two electrostatic mirrors (indicated in red), which are separated by a drift tube. Lens and steering electrodes as well as in-trap deflectors are included.}
	\label{Fig:MR-ToF_inventor}
\end{figure*}

A sectional view of the system is shown in Fig. \ref{Fig:MR-ToF_inventor}. It includes the electrostatic mirrors, lens and steering electrodes in front of the entrance mirror and behind exit mirror as well as in-trap deflector electrodes. The central drift tube serves as an in-trap potential lift.
\noindent
To maintain high flexibility, all in-vacuum components are combined into individual subassemblies that are placed on separate base plates, see Sec. \ref{Sec:Subassemblies} and \ref{Sec:Support_system}. All modules are in turn supported by a base plate ranging over the full length, which is referred to as the optical table in the following. Apart from electrically nonconducting materials and electrode supporting base plates made of aluminium and titanium, all in-vacuum components are made of stainless steel (material number 1.4404). For electrical insulating components, alumina ceramic and Macor\textsuperscript{\textregistered} glass ceramic are chosen, which allow bake-out temperatures up to $\SI{800}{\celsius}$. With the materials used, it is expected to reach a vacuum below $\SI{e-9}{\milli\bar}$ inside the MR-ToF MS during operation, reducing ion losses, especially for long storage times.

\subsection{Electrode assemblies}\label{Sec:Subassemblies}
Separated by the drift tube, the symmetrical arms of the MR-ToF mass spectrometer are formed by three subassemblies, which each serve individual purposes. This modular character offers the advantage of flexible assembly and easy modification for adaptations. Each subassembly is mounted onto its own base plate before it is fixed to the $\SI{1200}{\milli\meter}$ long, $\SI{90}{\milli\meter}$ wide and $\SI{4}{\milli\meter}$ thick optical table. Thus, it can be dismounted, modified and reintegrated without changing the remaining system. In the following, the three electrode modules are described in more detail, from the inside out.
\begin{figure*}[h]
	\centering
	\includegraphics[width = 1.0\textwidth]{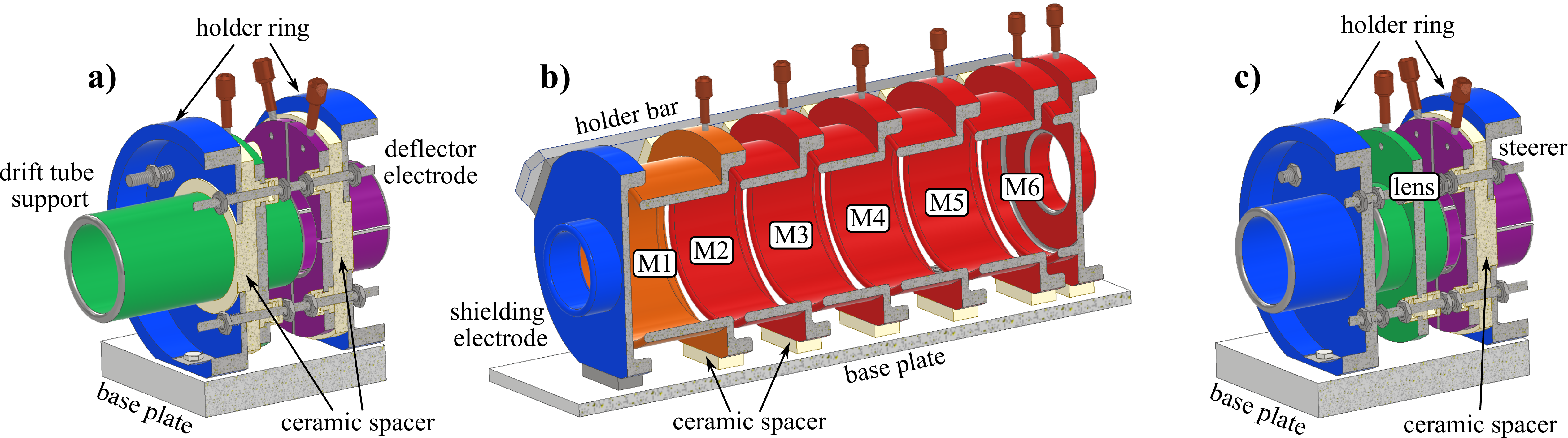}
	\caption{CAD representation of the individual electrode modules in cross-section view. The innermost electrode stack (a) includes a support for the drift tube and an in-trap deflector electrode for mass-selective ion deflection. Formed by a shielding electrode and six mirror electrodes, the electrostatic mirror (b) creates the trapping potential. At the injection and ejection sides, respectively, an einzel lens and an ion steerer allow for ion-beam adjustments (c). Both arms of the MR-ToF mass spectrometer are arranged with mirror symmetry.}
	\label{Fig:MR-ToF_modules}
\end{figure*}
\noindent
\\
The innermost electrode assembly includes a support electrode for the drift tube as well as an in-trap deflector electrode. The central drift tube can be plugged onto the support and is fixed to it by integrated screws. The in-trap deflector, which enables mass-selective ion deflection during the ion trapping phase, is formed by a quartered ring electrode which connects to the drift section. When applying an attractive or a repulsive potential to a deflector segment (or opposite potentials to opposing segments), ions can be kicked off their trajectory and are thus removed from the beam \cite{Fischer2018015114}. Both components, the drift tube support and the in-trap deflector, are supported by two holder rings from which they are insulated by ceramic spacers. In order to define the axial orientation, all components are mounted on four threaded rods. Their axial distance is specified and fixed by nuts on these rods (see Fig. \ref{Fig:MR-ToF_modules}a).\\
Following the innermost module outwards, the mirror-electrode stack forms the core of the MR-ToF system. A shielding electrode and six mirror electrodes are stacked with a distance of $\SI{5}{\milli\meter}$, electrically insulated by MACOR\textsuperscript{\textregistered} spacer plates and mounted on a common base plate. To keep the electrodes axially aligned, insulated holder bars fix the electrode orientation at two additional positions (see Fig. \ref{Fig:MR-ToF_modules}b).\\
At both ends of the MR-ToF MS, a lens and ion steering module completes the assembly. Similar to the in-trap module, presented above, a lens electrode and a quartered steering electrode, identical in construction to the in-trap deflector, are supported by four threaded rods and two mounting rings. One of the latter, together with the steering electrode, serves as grounded lens segment in front of and behind the lens electrode (see Fig. \ref{Fig:MR-ToF_modules}c). To protect the ceramics from mechanical stress during temperature fluctuations, the assemblies, which are secured by threaded rods, are spring-loaded. It allows movement of the electrodes and decouples the electrodes from the thermal expansion of the threaded rods. All electrodes are connected to standard SHV feedthroughs via in-vacuum Kapton cables and beryllium-copper connectors, which are plugged onto set screws integrated in the electrodes.

\subsection{In-vacuum support system}\label{Sec:Support_system}
The optical table that supports the complete MR-ToF system is made of titanium. It is reinforced by two additional $\SI{1200}{\milli\meter}$ long, $\SI{30}{\milli\meter}$ wide and $\SI{3}{\milli\meter}$ thick titanium plates to form a U-shaped profile. Originally, only one plate was planned but it was found that the plate bends too much under the weight of the electrodes. The material has been chosen because of the low thermal expansion coefficient, $\SI{8.35e-6}{\kelvin^{-1}}$ for temperatures between $\SI{20}{\celsius}$ and $\SI{100}{\celsius}$, compared to $\SI{16.5e-6}{\kelvin^{-1}}$ for 1.4404-type stainless steel and $\SI{23.03e-6}{\kelvin^{-1}}$ for aluminium \cite{Martienssen2005}. The lowest possible thermal expansion is in the interest of a stable potential profile and thus an improved mass resolving power for elongated measurement timescales. Consequently, it is advantageous to limit temperature fluctuations of the in-vacuum components and electrodes. They may most likely arise from variations of the temperature in the laboratory environment. Therefore, it is preferable to decouple the in-vacuum system as much as possible from the surrounding vacuum chamber. Hence, the titanium plate and, thus, the entire MR-ToF electrode system is designed to rest only on the conical tips of four set screws while its position is fixed by two additional screws. The set screws that define the support points are height-adjustable through threaded holes integrated in two dedicated mounting plates attached to the vacuum chamber.

\section{Current status}
\begin{figure}[H]
	\centering
	\includegraphics[width = 0.5\textwidth]{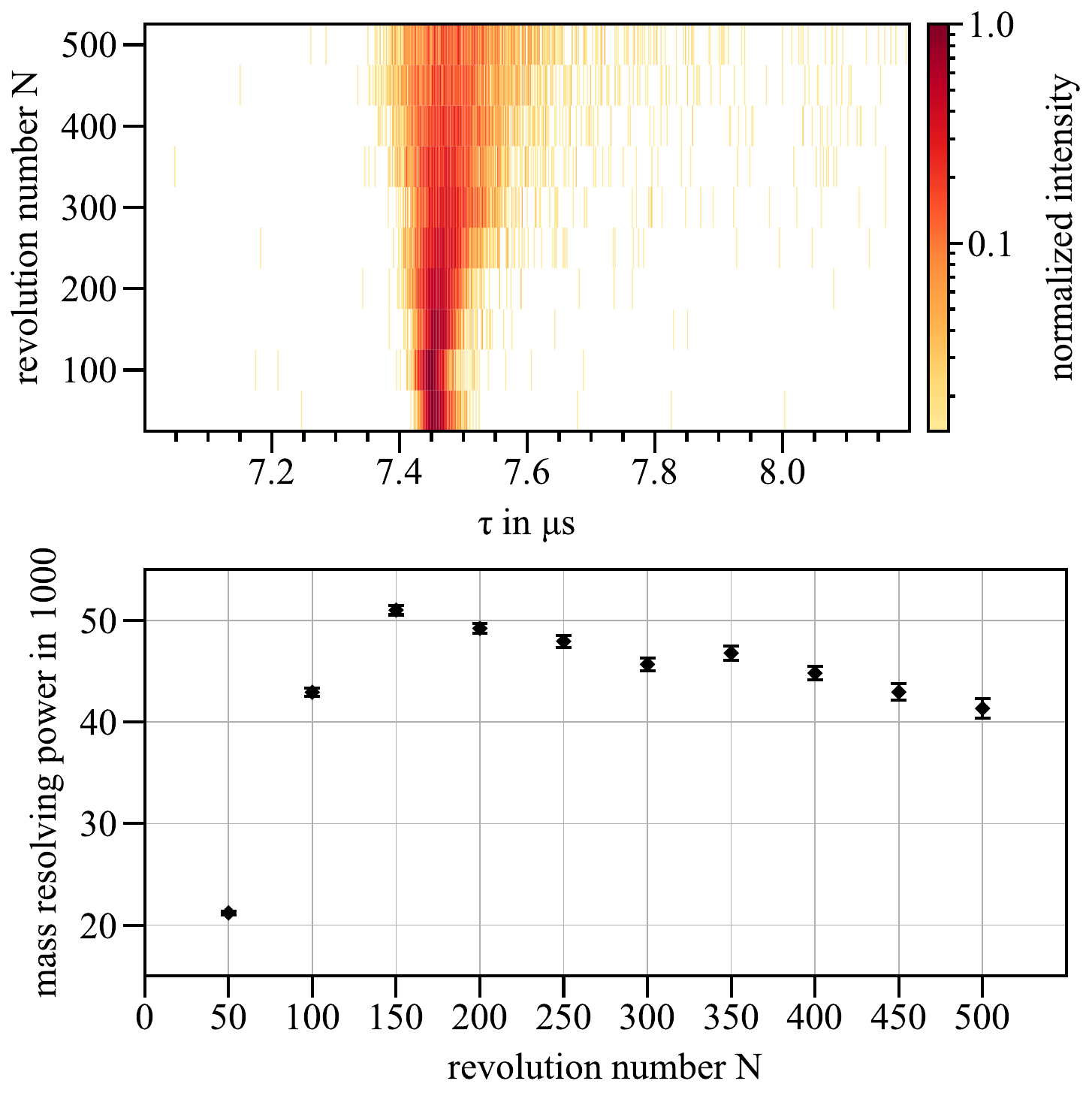}
	\caption{Ion intensity of $^{40}$Ar$^+$ ions measured as a function of the ToF after the ejection from the MR-ToF MS $\tau$ and the number of revolutions $N$ for which the ions are trapped in the MR-ToF MS for (top). Evolution of the mass resolving power as a function of $N$ (bottom).}
	\label{Fig:Argon-operation}
\end{figure}

The PUMA offline ion source setup as shown in Fig. \ref{Fig:Offline_ion_source} has been assembled from the ion source to the Paul trap. The electron impact ionisation source has been commissioned with different gases, such as air, hydrogen, argon, neon and xenon. With an electron energy of about $\SI{100}{\electronvolt}$, multiply-charged ions can be generated. For the commissioning of the MR-ToF MS, using argon as target material, a continuous ion beam of approximately $\SI{10}{\pico\ampere}$ is extracted and chopped by pulsing a pair of deflector electrodes included in the ion source with a $\SI{200}{\nano\second}$ gate from a deflection to a transmission state. This results in ion packets with a FWHM of the ToF distribution of about $\SI{250}{\nano\second}$, measured with a ToF detector behind the MR-ToF MS. Due to the largest abundance, $^{40}$Ar$^+$ was selected to be used for first test measurements. Starting from the potential configuration determined in Sec. \ref{Sec:Potential_determination} and compensating possible misalignments by manually tuning the mirror electrodes and the in-trap lift voltage, the ToF focus could be set on the detector behind the device after trapping the ions for about 100 - 150 revolutions. For longer storing times and the given voltage applied to the in-trap lift electrode, the ToF signal is defocused again. This is shown in Fig. \ref{Fig:Argon-operation} together with a plot of the mass resolving power as a function of the number of revolutions that the $^{40}$Ar$^+$ ions are trapped in the MR-ToF MS for. With a signal width of $\Delta t = \SI{34}{\nano\second}$ (FWHM), the highest mass resolving power of $R = m/\Delta m = t/2\Delta t \sim 50,000$ is reached after 150 revolutions.  
\noindent
\begin{figure*}[h]
	\centering
	\includegraphics[width = 1.0\textwidth]{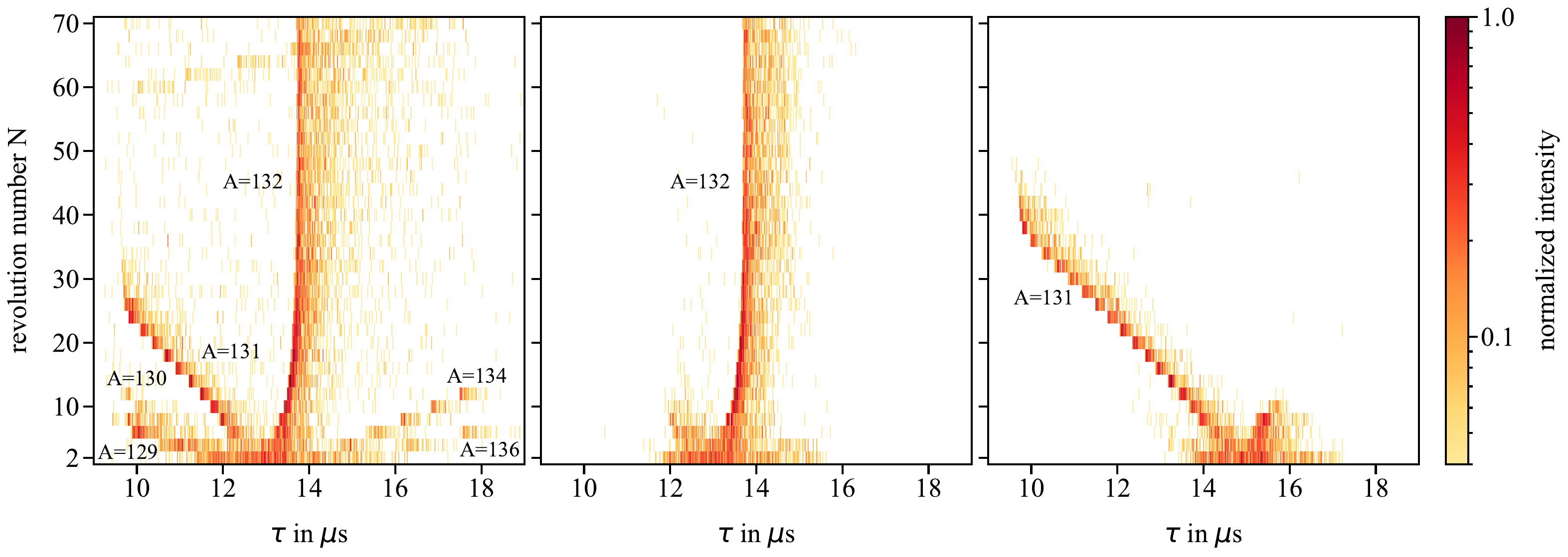}
	\caption{Ion intensity of Xe$^+$ ions measured as a function of the ToF after the ejection from the MR-ToF MS $\tau$ and the number of revolutions $N$ for which the ions are trapped in the MR-ToF MS for. All three plots show the same measurement, where once the deflector is switched off (left), once the deflector is switched on and synchronised to A=132 (centre) and once synchronised to A=131 (right). In case of the right plot, the capture time is shifted by $\SI{2}{\micro\second}$ to keep the $^{132}$Xe$^+$ ions in the lift for a larger number of revolutions.}
	\label{Fig:Xenon-operation}
\end{figure*}
\noindent
\\
In order to demonstrate the capability for isotopic purification, the in-trap deflector electrodes are commissioned using xenon as target gas, which has nine natural isotopes. A fast high voltage switch (Stahl-Electronics, HS-2000) is used to switch the potential applied to two opposite segments of the quadruple segmented deflector electrode between \SI{0}{\volt} and \SI{\pm100}{\volt}, respectively. Thus, the in-trap deflector can be switched between a deflecting (\SI{200}{\volt} potential difference) and a transmission state (\SI{0}{\volt} potential difference). The rising and falling times are typically below \SI{500}{\nano\second} and therefore enable undisturbed deflection of ion species separated by a few \SI{}{\micro\second}. By synchronising the switch to the revolution period of a desired ion mass-over-charge ratio, the deflector can be used for a mass-selective transversal ion ejection \cite{Fischer2018015114}.
The left-hand plot in Fig. \ref{Fig:Xenon-operation} shows the measured ion intensity of singly-charged xenon isotopes as a function of the time-of-flight $\tau$ after the ejection from the MR-ToF MS (abscissa) and the number of revolutions the ions are trapped in the MR-ToF MS for (ordinate). Here, the deflector is not active. Below $\tau = \SI{10}{\micro\second}$ (above $\tau = \SI{18}{\micro\second}$) ions with a smaller (larger) mass-to-charge ratio are not located in the in-trap lift electrode during switching, resulting in them not being ejected from the MR-ToF MS. The centre plot shows the same measurement but with the deflector activated using a duty cycle of 67\%. The latter refers to the ratio between the length of the deflection state of $\SI{27}{\micro\second}$ and the revolution period of $\SI{40.5}{\micro\second}$ for A/z=132 ions, where A is the mass number and z the charge state. In this case, the transmission window of the deflector is synchronised and centred on mass A/z=132. In the right-hand plot, the measurement is repeated with the transmission window synchronised and centred to mass A/z=131, allowing to isolate $^{131}$Xe ions. To increase the number of revolutions for which $^{131}$Xe ions can still be ejected, the capture time was decreased by $\SI{2}{\micro\second}$, which is the time between the ion extraction from the source and the switching of the in-trap lift. This results in a shift of $\tau$ by $\SI{2}{\micro\second}$ to a longer ToF.

\section{Conclusion and outlook}
In the framework of the PUMA project, a new MR-ToF-MS has been developed for the PUMA offline ion source setup, allowing for fast isotopic ion beam purification. To increase the transmission of the device and, thus, decreasing the accumulation time of $10^5$ ions in an RFQ ion trap, which is mandatory for PUMA, SIMION\textsuperscript{\textregistered} simulations have been performed modifying the electrode geometry of an MR-ToF MS design from the University of Greifswald. This has predominantly led to an increase in the inner diameters and the corresponding adjustment of the individual electrode lengths. As a result, the transmission could be increased by a factor of about eight, as shown by benchmarking simulations with realistic mirror potential profiles. To facilitate the assembly and allow for later modifications, the design is based on several electrode sub-assemblies. The setup can be used to capture the ions of interest and separate them from unwanted ion species. Furthermore, focus and steering electrodes are included in the design on both the injection and the ejection sides of the device. 
\\
The MR-ToF MS has been commissioned, using a chopped argon ion beam, extracted from a commercial electron impact ionisation source. First mirror tuning attempts have led so far to a mass resolving power of 50,000, trapping Ar$^+$ ions for 150 revolutions. Furthermore, the in-trap deflector has been commissioned using Xe$^+$ ions. For the examples $^{131}$Xe$^+$ and $^{132}$Xe$^+$, it could be demonstrated that the device is able to successfully isolate isotopes of interest from a stable ion spectrum.\\
Already during the simulation phase, an increasing number of institutes expressed their interest in an MR-ToF device. As a result, the Darmstadt's MR-ToF (DA's MR-ToF) Collaboration has been formed to distribute the manufacturing and acquisition of the different parts among seven institutions. These collaboration partners are the Technical University of Darmstadt, the University of Greifswald, the University of Groningen, the University of Manchester, the Massachusetts Institute of Technology, the Johannes Gutenberg University Mainz and the University of Innsbruck.

\section*{Acknowledgements}
We gratefully acknowledge the discussions with the staff of the mechanical workshop at TU Darmstadt and thank the contributing institutes and workshops for their support. The PUMA project is funded by the European Research Council through the ERC grant PUMA-726276. J.F., C.K., A.O., A.S. and M.S. appreciate the support from the Alexander von Humboldt Foundation.

\printbibliography 
\end{document}